\documentclass{aa}

\usepackage{graphicx}
\usepackage{txfonts}
\begin{document}

	\titlerunning{High time resolution broad-band polarimetry}
	\title{High time resolution broad-band polarimetry:}

	\subtitle{technique, calibration and standards}

	\author{
		V. Breus\inst{1}
		S. V. Kolesnikov \inst{1,2}
		and I. L. Andronov \inst{1}
	}

   \institute{
			Department of Mathematics, Physics and Astronomy, Odessa National Maritime University, Mechnikova 34, UA-65029 Odessa, Ukraine
			\email{vitaly.breus@gmail.com}
		\and
			Astronomical Observatory, Odessa National University, Marazliyevskaya 1b, UA-65014 Odessa, Ukraine
	}

   \date{Received Otober 19, 2021; accepted December 31, 2021}

 
  \abstract
   { Regular large-scale polarimetric observations in Crimean astrophysical observatory began in the early 1960s. In 2002 - 2017 the single-channel aperture photometer-polarimeter with a quarter-wave plate at the 2.6-m Shajn mirror telescope (SMT) was used. We accumulated a large homogeneous data set of polarimetric observations of different types of objects that are to be published separately. }
   { Correct polarimetric data processing requires high polarization standards and zero-polarization stars. We aim to improve the data reduction and calibration process to obtain further results with highest possible accuracy.}
   { High time resolution broad-band (WR, R, V, B, U) polarization observations are made of 98 known standard stars (527 time series with total duration about 184 hours). }
   { We determined values of linear and circular polarization for 98 nearby Northern bright stars. This catalogue is not compilative, but obtained using the same instrument and technique during large time interval. It will be used for our future research and it may be used by other authors. We implemented the least squares approach for determination of the Stokes parameters. It allowed us to obtain results with the accuracy better then we obtained using previously used methods. We report suspicious or variable stars that are not suitable as standards for high precision polarimetry. }
   {}


   \maketitle
%

\section{Introduction}
Polarization is an important property of light. It is the subject of observation, interpretation and simulation for multiple decades. Polarization has been regularly detected in scattered light. Polarimetric characteristics of the scattered radiation contain valuable information about such important properties of particles as their size, morphology, and chemical composition (Mishchenko et al. \citeyear{2010Mishchenko}). Circularly polarized light could be emitted from the sources with a strong magnetic field (e.g. magnetic cataclysmic binary systems - intermediate polars and polars, magnetic white dwarfs, accreting regions, neutron stars) and its properties are linked to the internal geometry of a source of radiation. As a consequence, polarimetry complements photometric and spectroscopic studies of sources of radiation and has made possible many astrophysical discoveries.

Correct polarimetric data processing is a particularly crucial task. It consists not only of the elimination of the sky background that may be linearly polarized, but also the subtraction of instrumental polarization, polarization scale calibration and determination of the zero-point of polarization angle. These tasks require high polarization standards and zero-polarization stars.
During decades of polarimetric observations different lists or catalogs of polarization standards were created. The interstellar polarization was first studied by \citet{1959VeGoe...7..199B}. Polarimetric observations of nearby stars were made using rotatable tube telescopes e.g. by \citet{1966ZA.....64..269A}, \citet{1968ApJ...154..115S}, \citet{1968PASP...80..162W}, \citet{1975ApJ...196..261S}. \citet{1977A&AS...30..213P} measured polarization of 77 stars within 25 pc from the Sun and compared the data with obtained using rotatable tube telescopes. \citet{1993A&AS..101..551L} compiled a catalogue of optical polarization measurements for 1000 stars closer then 50 pcs from the Sun.
\citet{2000AJ....119..923H} published an agglomeration of stellar polarization catalogs with data for 9286 stars. Recently, \citet{2016MNRAS.455.1607C} published the results of observations of 50 nearby Southern bright stars with high accuracy.

Regular large-scale polarimetric observations in Crimean astrophysical observatory began in the early 1960s. A brief history and descriptions of the equipment were presented in the book by \citet{2010Mishchenko}, pp. 106-112. For years, methods of data processing and software were elaborated (\citet{1972IzKry..45...90S}; \citet{1981IzKry..63..118E}; \citet{1998KPCB...14..359S}; \citet{2001IzKry..97...91S}; \citet{2007A&AT...26..241B}; \citet{2007OAP....20...32B}; \citet{2016OAP....29...74K}; \citet{2019JPhSt..23.3901K}; \citet{2019JPhSt..23.4901K}) and many important results were published (\citet{2008LPICo1405.8393R}; \citet{2013JPhSt..17.3901B}).
At the end of 2017 this project was finished for reasons unrelated to astrophysics. This allowed us to reconsider our approach to data analysis.

The paper is organized as follows: in the next Section we describe the data and the instruments used, and Section~\ref{sec:Dataprocessing} presents the methods used for data processing and derivation of the Stokes parameters. In Sections~\ref{sec:LPCP} and \ref{sec:Discussion} we describe results obtained from circular and linear polarization. We discuss our results in the last Section.


\section{Observations}
\label{sec:Observations}

\begin{figure}
    \includegraphics[width=0.75\hsize]{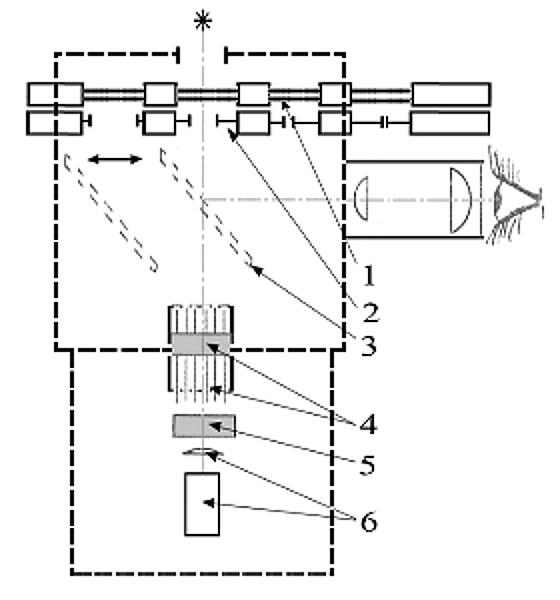}
        \caption{Block diagram of the single-channel photometer-polarimeter used in this work.}
        \label{fig:optscheme}
\end{figure}

The observations were obtained using single-channel aperture photometer-polarimeter installed on the Cassegrain focus of the 2.6-m Shajn mirror telescope (SMT) at the Crimean astrophysical observatory (Ukraine). A rotating strained plexiglass quarter-wave plate calculated by \citet{1986KFNT....2...82K} and developed by Samoilov \& Samoilov (2015) was used as a circular polarization modulator. After the tests under laboratory conditions it turned out that optical efficiency of this plate is similar to quartz plates, but plexiglass is less sensitive to temperature changes. The retarder plate is continuously rotating at the rate of 33 rps. It allows us to measure circular and linear polarization simultaneously. Optical system of the polatimeter is presented at the Fig.~\ref{fig:optscheme}. It consists of the:

\begin{enumerate}
	\item filter wheel
	\item aperture
	\item mobile mirror
	\item the quarter-wave plate rotating at the rate of 33 rps
	\item Glan prism (analyzer)
	\item photomultiplier tube with the Fabry lens
\end{enumerate}
Detailed description of the instrument was presented by \citet{2016OAP....29...74K}.

The possibility of linear to circular cross-talk in the instrument was studied in laboratory conditions: the maximal value was found to be 1.5

We used mostly WR (Wide-R, 550-750 nm) and Johnson V and R filters. Few measurements were made using B and U flters. The typical object exposures are 1-10 seconds that depends on the target's brightness and rate of variability. It allows us to study fast changes of each parameter or obtain smaller time resolution but higher accuracy by averaging.
The whole data set contains about 1000 time series of different objects in 105 sets of observations between September 2002 and November 2017. Hereafter the set of observations is the period between the installation and removing of the polarimeter from the telescope (usually 1 to 5 consecutive nights).


\section{Data processing}
\label{sec:Dataprocessing}

The intensity of light transmitted through a retarder with a quarter wave plate is described by the formula~\ref{eq:intens}, where $I$, $Q$, $U$, $V$ - Stokes parameters, $\psi$ is the angle of rotation of the retarder axis relative to the analyzer principal plane (Serkowski \citeyear{1974psns.coll..135S}).

\begin{equation}
I = \frac{1}{2} \left(I \pm \frac{1}{2}Q \pm \frac{1}{2}Q\cos4\psi \pm \frac{1}{2}U\sin4\psi \mp V\sin2\psi\right)
\label{eq:intens}
\end{equation}

After angular integration of this equation for eight pulse counters (22.5 degrees each) and taking into account repeating after 180 degrees, 8 equations could be obtained (see Kolesnikov et al. \citeyear{2016OAP....29...74K}). These data is accumulated in 8 counters of the controller and stored for each exposure.
To eliminate the background, series of target were alternated with series of background and series of comparison star to obtain photometry.

The background was fit by polynomials and subtracted from the data for each of 8 channels independently. Due to the fast rotation of the phase plate, we can exclude influence of majority of atmospheric phenomena.
Twilight sky is characterized by rapid change of the background level, so the background series were made more frequently and a proper fit required high order polynomials. Each normal night with good atmospheric conditions shows slow background variations and absence of background polarization, so most of backgrounds were fitted using 0 or 1st order polynomials. Clouds or haze sometimes were linearly polarized, sometimes were not. We noticed that in the second case, partial haze or clouds were obviously harmful for photometry, but did not change the polarization, only increased dispersion was observed. So, if continuous series of particular target is highly important and the data loss is critical, it is possible to use data obtained using this type of instrument even through significant cloudiness. However, we excluded all suspicious points from the analysis of standards.

Contrary to classic algorithm used earlier (\citet{2007A&AT...26..241B}; \citet{2016OAP....29...74K}), based on determination of Stokes parameters from linear combinations of 8 equations, we used least squares approach. This way we get some benefits:

\begin{enumerate}
\item we can obtain all four Stokes parameters $I$, $Q$, $U$, $V$ and their accuracies using trivial formulas
\item we do not need to determine a position angle of circular polarization diagram using objects with high-amplitude variability of circular polarization, to obtain circular polarization degree from $CP1$ and $CP2$
\item we stop neglecting influence of non-zero $Q$ on $I$, valid only for signal with low values of $Q$
\item the accuracy of the results obtained using this approach is obviously better.
\end{enumerate}

Also we may exclude the influence of artificial noise or electrical interference in conductors by excluding up to 4 erroneous channels (with decrease of the results accuracy).
As interference happened irregularly and could not be predicted, detection of interference was implemented using complicated algorithm based on analysis of moving average in each of 8 channels.
Normalized Stokes parameters  were calculated as follows:

\begin{equation}
q = \frac{Q}{I},
u = \frac{U}{I},
v = \frac{V}{I}
\label{eq:normalized}
\end{equation}

The total linear polarization
\begin{equation}
p = \sqrt[]{q^2 + u^2}
\label{eq:plinear}
\end{equation}

Throughout this work, the normalized Stokes parameters $q$, $u$, $v$ and linear polarization degree $p$ are given in per cents, position angle $\theta$ is given in degrees. All measurements of $\theta$ are referred to equatorial coordinates.

Measurements of zero-polarization stars are used for computation of the instrumental polarization. Classic approach implies usage of the measurements obtained each night before and/or after the observations, in the same filter, similarly to the flat field and dark frames in CCD photometry.
Usually the mean values of normalized Stokes parameters over the observations set were used (one set is a time interval that consist of a few days up to one week of continuous observations without dismantling the equipment).
We found that these mean values are not statistically different from set to set. We noticed different behaviour, which clearly separated the period before replacing the reflective layer of the main mirror of the telescope and after it. It was originally detected with the dependency of linear polarization degree on time, but later determined for normalized Stokes parameters q and u separately. For example, at the first (flat) interval, $q = 0.026(10)\%$, $u = 0.036(12)\%$, so the magnitude of the instrumental polarization is not negligible. The difference of fits of instrumental polarization obtained for WR, R, V and B filters is not statistically significant, so we used all measurements together.
Thus the instrumental polarization was determined in the form of piecewise continuous functions from the observations of about 65 non-polarized stars for the whole data interval. Subtracting these values from the normalized Stokes parameters we eliminate the instrumental linear polarization.
In each set, few different highly polarization standards were observed for calibration of position angle zero-point. We checked that the difference between position angle in our instrumental system and the catalog value for the corresponding star (\citet{2000AJ....119..923H}; \citet{1993A&AS..101..551L}) was constant during each set of observations for absolute majority of standards. Contrary to the classic approach, where the position angle was determined for each set of observations, taking into account statistically insignificant difference between many consecutive sets, we decided to determine the position angle for larger time intervals. Weighted mean values of the difference were calculated.  Here and later we used the weight inversely proportional to the square of the error estimate of the value. Summarizing, about one third of all data set time interval had the same position angle value, one third had another value, and the rest of the data (the earliest years) show position angle changes from set to set due to changes of the device. We have rotated the axes $Q$, $U$ on these angles to align our measurements to equatorial coordinates.
Thus, the weighted mean values of the instrumental linear polarization and position angle were determined using significantly larger number of observations (few sets instead of one) and obviously are more accurate than determined earlier by few measurements of standard stars during one or two consequent sets.
After the correction for background and instrumental polarization, the intensity $I$ and normalized Stokes parameters $q$, $u$, $v$ with its error estimates are stored in the form of time series and are subject for further analysis.
For the constant objects (like polarization standards and zero-polarization stars) we used weighted mean of Stokes parameters for each run. The errors were computed as the standard errors of the weighted mean values.

\begin{equation}
\begin{aligned}
\theta = 0.5 \arctan \frac{u}{q} \\
\sigma\left(p\right) = \sqrt[]{\sigma\left(q\right)^2 + \sigma\left(u\right)^2} \\
\sigma\left(\theta\right) = \arctan \frac{ 0.5 \sigma\left(p\right) }{p}
\end{aligned}
\label{eq:ptheta}
\end{equation}

The final values of each Stokes parameter were obtained as follows: the weighted mean value and its standard error were calculated for each object during the first iteration. Then we excluded all runs, that deviates from the mean value for more then $2\sigma$ and repeated the calculation of weighted mean and the error. The linear polarization degree, the position angle and their error estimates were calculated from the final values of Q and U using Formula~\ref{eq:ptheta}.

These results are shown as normalized Stokes parameters in the Table~\ref{tab:Table1part1}.

\begin{table*}
    \centering
    \caption{Results of standard polarized and unpolarized stars. From the left to the right are given: object name, band, number of different nights, total number of exposures, sum of exposure durations in minutes, normalized Stokes parameters q, u, v, the linear polarization degree in percents and the position angle referred to equatorial coordinates with their error estimates.}
    \label{tab:Table1part1}
    \begin{tabular}{lccrrrrrrrrrrrr}
		\hline\hline
Object & Filter & N & NP & $\delta$T, min & q & $\sigma$q & u & $\sigma$u & v & $\sigma$v & p & $\sigma$p & $\theta$ & $\sigma\theta$\\
\hline
BD+332642 & V & 1 & 512 & 54 & 0.168 & 0.045 & 0.059 & 0.045 & 0.001 & 0.020 & 0.178 & 0.063 & 99.6 & 10.0\\
BD+332642 & WR & 6 & 1708 & 161 & -0.039 & 0.018 & 0.089 & 0.018 & -0.001 & 0.008 & 0.097 & 0.025 & 146.9 & 7.3\\
BD+64 106 & V & 2 & 632 & 51 & 5.415 & 0.044 & 1.382 & 0.043 & 0.028 & 0.021 & 5.589 & 0.061 & 97.2 & 0.3\\
BD+64 106 & WR & 2 & 352 & 26 & 5.261 & 0.040 & 1.550 & 0.039 & 0.008 & 0.018 & 5.485 & 0.055 & 98.2 & 0.3\\
BD284211 & V & 4 & 1069 & 80 & 0.605 & 0.026 & -0.298 & 0.026 & -0.014 & 0.012 & 0.674 & 0.037 & 76.9 & 1.6\\
BD284211 & WR & 4 & 734 & 94 & -0.005 & 0.027 & 0.050 & 0.027 & -0.041 & 0.012 & 0.050 & 0.037 & 138.1 & 20.6\\
G191B2B & V & 1 & 600 & 44 & -0.303 & 0.088 & 0.336 & 0.088 & 0.217 & 0.041 & 0.453 & 0.125 & 156.0 & 7.9\\
G191B2B & WR & 3 & 628 & 89 & 0.025 & 0.028 & 0.244 & 0.028 & 0.003 & 0.013 & 0.246 & 0.040 & 132.1 & 4.6\\
HD102870 & R & 1 & 512 & 36 & -0.023 & 0.011 & -0.063 & 0.011 & 0.049 & 0.005 & 0.067 & 0.016 & 35.2 & 6.9\\
HD102870 & WR & 6 & 1102 & 90 & -0.040 & 0.002 & 0.012 & 0.002 & -0.013 & 0.001 & 0.042 & 0.003 & 171.9 & 1.9\\
HD103095 & WR & 1 & 320 & 24 & -0.066 & 0.017 & 0.055 & 0.017 & 0.030 & 0.008 & 0.086 & 0.024 & 160.1 & 8.1\\
HD10476 & R & 3 & 1600 & 114 & -0.015 & 0.005 & -0.073 & 0.005 & 0.030 & 0.002 & 0.075 & 0.007 & 39.1 & 2.8\\
HD10476 & WR & 8 & 1344 & 84 & 0.021 & 0.003 & 0.029 & 0.003 & 0.054 & 0.001 & 0.036 & 0.004 & 116.9 & 3.4\\
HD10898 & R & 1 & 283 & 19 & 3.454 & 0.047 & 0.316 & 0.047 & 0.070 & 0.022 & 3.468 & 0.067 & 92.6 & 0.6\\
HD10898 & V & 1 & 703 & 54 & 4.244 & 0.022 & 0.374 & 0.021 & -0.002 & 0.010 & 4.260 & 0.030 & 92.5 & 0.2\\
HD10898 & WR & 1 & 80 & 14 & 3.561 & 0.049 & 0.755 & 0.048 & 0.000 & 0.022 & 3.640 & 0.068 & 96.0 & 0.5\\
HD109358 & WR & 7 & 1601 & 133 & -0.003 & 0.002 & 0.024 & 0.002 & 0.046 & 0.001 & 0.024 & 0.003 & 138.2 & 3.8\\
HD110897 & WR & 2 & 494 & 35 & -0.018 & 0.008 & 0.011 & 0.008 & -0.011 & 0.004 & 0.021 & 0.011 & 164.4 & 15.5\\
HD111395 & R & 1 & 84 & 18 & -0.064 & 0.072 & -0.229 & 0.073 & -0.057 & 0.033 & 0.238 & 0.102 & 37.2 & 12.1\\
HD114710 & R & 1 & 512 & 36 & -0.061 & 0.014 & -0.027 & 0.014 & 0.020 & 0.006 & 0.066 & 0.019 & 12.2 & 8.3\\
HD114710 & WR & 12 & 2767 & 208 & -0.053 & 0.001 & 0.007 & 0.001 & 0.000 & 0.001 & 0.053 & 0.002 & 176.1 & 1.1\\
HD125184 & WR & 2 & 1280 & 92 & -0.151 & 0.008 & -0.037 & 0.008 & 0.035 & 0.004 & 0.156 & 0.011 & 6.8 & 2.1\\
HD126660 & WR & 5 & 966 & 61 & -0.007 & 0.002 & -0.012 & 0.002 & 0.011 & 0.001 & 0.014 & 0.003 & 29.8 & 6.4\\
HD127665 & WR & 3 & 576 & 57 & -0.035 & 0.009 & 0.348 & 0.009 & 0.016 & 0.004 & 0.350 & 0.012 & 137.9 & 1.0\\
HD127762 & WR & 9 & 1216 & 96 & -0.020 & 0.001 & -0.044 & 0.001 & 0.016 & 0.001 & 0.049 & 0.002 & 32.6 & 1.2\\
HD14069 & R & 1 & 640 & 46 & -0.031 & 0.011 & -0.082 & 0.011 & 0.147 & 0.005 & 0.087 & 0.016 & 34.6 & 5.1\\
HD14069 & V & 6 & 2767 & 171 & 0.295 & 0.014 & 0.153 & 0.014 & -0.072 & 0.006 & 0.332 & 0.020 & 103.7 & 1.7\\
HD14069 & WR & 7 & 2344 & 127 & 0.052 & 0.004 & -0.005 & 0.004 & 0.078 & 0.002 & 0.052 & 0.006 & 87.3 & 3.2\\
HD142373 & WR & 2 & 192 & 14 & 0.026 & 0.006 & -0.046 & 0.006 & 0.007 & 0.003 & 0.052 & 0.008 & 59.7 & 4.4\\
HD144287 & B & 1 & 62 & 12 & -0.153 & 0.251 & 0.042 & 0.250 & -0.047 & 0.116 & 0.158 & 0.354 & 172.3 & 48.2\\
HD144287 & R & 1 & 576 & 46 & 0.011 & 0.030 & 0.111 & 0.030 & 0.060 & 0.014 & 0.111 & 0.043 & 132.3 & 10.9\\
HD144287 & V & 3 & 632 & 65 & 0.013 & 0.016 & -0.003 & 0.016 & 0.046 & 0.007 & 0.014 & 0.022 & 84.1 & 39.1\\
HD144287 & WR & 8 & 1912 & 165 & 0.063 & 0.008 & 0.021 & 0.008 & 0.030 & 0.003 & 0.067 & 0.011 & 99.3 & 4.6\\
HD14433 & B & 1 & 78 & 20 & 2.144 & 1.061 & 2.961 & 1.060 & 0.082 & 0.497 & 3.656 & 1.500 & 117.0 & 11.6\\
HD14433 & V & 2 & 87 & 19 & 2.615 & 0.099 & 2.801 & 0.099 & -0.069 & 0.046 & 3.832 & 0.139 & 113.5 & 1.0\\
HD14433 & WR & 3 & 396 & 35 & 2.321 & 0.062 & 2.587 & 0.063 & -0.041 & 0.029 & 3.475 & 0.088 & 114.1 & 0.7\\
HD144579 & R & 1 & 160 & 12 & -0.018 & 0.018 & -0.093 & 0.018 & 0.082 & 0.008 & 0.095 & 0.026 & 39.5 & 7.7\\
HD144579 & V & 1 & 60 & 10 & -0.057 & 0.063 & 0.059 & 0.063 & -0.036 & 0.029 & 0.082 & 0.089 & 157.0 & 28.5\\
HD144579 & WR & 2 & 120 & 20 & -0.226 & 0.046 & -0.057 & 0.046 & -0.023 & 0.021 & 0.233 & 0.065 & 7.1 & 8.0\\
HD146233 & V & 1 & 128 & 9 & 0.017 & 0.024 & 0.063 & 0.024 & -0.088 & 0.011 & 0.065 & 0.034 & 127.5 & 14.7\\
HD146233 & WR & 2 & 504 & 102 & 0.271 & 0.006 & -0.030 & 0.006 & 0.022 & 0.003 & 0.273 & 0.009 & 86.8 & 0.9\\
HD15089 & WR & 3 & 640 & 70 & -0.258 & 0.011 & -0.002 & 0.011 & -0.080 & 0.005 & 0.258 & 0.015 & 0.2 & 1.7\\
HD154345 & B & 1 & 80 & 15 & 0.034 & 0.049 & -0.043 & 0.049 & 0.035 & 0.023 & 0.055 & 0.069 & 63.9 & 32.2\\
HD154345 & R & 3 & 640 & 72 & -0.107 & 0.032 & 0.070 & 0.032 & -0.072 & 0.015 & 0.128 & 0.046 & 163.3 & 10.1\\
HD154345 & V & 6 & 700 & 136 & 0.008 & 0.017 & 0.015 & 0.017 & -0.034 & 0.008 & 0.017 & 0.024 & 120.5 & 34.6\\
HD154345 & WR & 15 & 2415 & 300 & 0.044 & 0.005 & -0.060 & 0.005 & 0.000 & 0.002 & 0.074 & 0.007 & 63.1 & 2.7\\
HD154892 & R & 1 & 640 & 47 & -0.072 & 0.044 & 0.005 & 0.044 & 0.077 & 0.020 & 0.072 & 0.062 & 178.0 & 23.3\\
HD154892 & V & 2 & 640 & 46 & 0.085 & 0.018 & -0.064 & 0.018 & 0.075 & 0.008 & 0.106 & 0.025 & 71.6 & 6.8\\
HD154892 & WR & 8 & 1589 & 117 & 0.050 & 0.008 & -0.036 & 0.008 & -0.007 & 0.004 & 0.062 & 0.011 & 71.9 & 5.2\\
HD155528 & V & 1 & 511 & 37 & 5.210 & 0.047 & 0.278 & 0.047 & 0.105 & 0.022 & 5.217 & 0.067 & 91.5 & 0.4\\
HD155528 & WR & 2 & 96 & 6 & 4.567 & 0.059 & -0.853 & 0.060 & 0.092 & 0.028 & 4.646 & 0.084 & 84.7 & 0.5\\
HD157214 & V & 2 & 256 & 17 & 0.034 & 0.010 & -0.078 & 0.010 & -0.072 & 0.004 & 0.085 & 0.014 & 56.6 & 4.6\\
HD157214 & WR & 4 & 320 & 22 & 0.049 & 0.006 & 0.117 & 0.006 & 0.004 & 0.003 & 0.127 & 0.008 & 123.6 & 1.8\\
HD165908 & V & 2 & 192 & 14 & -0.025 & 0.008 & -0.057 & 0.008 & -0.026 & 0.003 & 0.062 & 0.011 & 33.0 & 4.9\\
HD165908 & WR & 9 & 1376 & 111 & 0.001 & 0.003 & 0.019 & 0.003 & 0.005 & 0.001 & 0.019 & 0.004 & 133.5 & 5.8\\
HD182572 & WR & 3 & 800 & 64 & 0.111 & 0.004 & 0.010 & 0.004 & -0.017 & 0.002 & 0.112 & 0.006 & 92.5 & 1.5\\
HD185395 & WR & 3 & 448 & 33 & 0.053 & 0.006 & 0.037 & 0.006 & -0.036 & 0.003 & 0.065 & 0.009 & 107.4 & 3.8\\
HD187929 & WR & 3 & 352 & 31 & 1.370 & 0.005 & -0.074 & 0.005 & 0.019 & 0.002 & 1.372 & 0.007 & 88.5 & 0.1\\
HD18803 & R & 2 & 1019 & 86 & 0.001 & 0.015 & -0.023 & 0.015 & 0.019 & 0.007 & 0.023 & 0.021 & 45.6 & 24.7\\
HD18803 & WR & 6 & 965 & 176 & 0.068 & 0.010 & 0.062 & 0.010 & -0.009 & 0.004 & 0.092 & 0.014 & 111.3 & 4.3\\
            \hline
    \end{tabular}
\end{table*}

\begin{table*}
    \centering
    \caption{Continuation of the Table~\ref{tab:Table1part1}}
    \label{tab:Table1part2}
    \begin{tabular}{lccrrrrrrrrrrrr}
		\hline\hline
Object & Filter & N & NP & $\delta$T, min & q & $\sigma$q & u & $\sigma$u & v & $\sigma$v & p & $\sigma$p & $\theta$ & $\sigma\theta$\\
\hline
HD188326 & B & 3 & 796 & 80 & 0.295 & 0.027 & 0.112 & 0.027 & 0.144 & 0.012 & 0.315 & 0.038 & 100.4 & 3.5\\
HD188326 & R & 4 & 1864 & 162 & 0.013 & 0.012 & -0.104 & 0.012 & 0.023 & 0.005 & 0.105 & 0.017 & 48.5 & 4.6\\
HD188326 & V & 4 & 554 & 65 & 0.114 & 0.020 & -0.079 & 0.020 & 0.009 & 0.009 & 0.139 & 0.028 & 72.6 & 5.8\\
HD188326 & WR & 8 & 835 & 114 & -0.001 & 0.009 & -0.045 & 0.009 & 0.017 & 0.004 & 0.045 & 0.013 & 44.1 & 8.2\\
HD188512 & V & 1 & 64 & 4 & -0.045 & 0.008 & -0.010 & 0.008 & -0.040 & 0.003 & 0.046 & 0.011 & 6.2 & 6.7\\
HD188512 & WR & 5 & 577 & 52 & 0.013 & 0.003 & -0.022 & 0.003 & -0.005 & 0.002 & 0.026 & 0.005 & 60.2 & 5.4\\
HD18883 & WR & 1 & 121 & 12 & -0.013 & 0.036 & -0.079 & 0.036 & -0.039 & 0.016 & 0.080 & 0.050 & 40.5 & 17.4\\
HD190406 & R & 1 & 576 & 41 & 0.047 & 0.022 & -0.033 & 0.022 & 0.145 & 0.010 & 0.057 & 0.031 & 72.6 & 15.3\\
HD190406 & WR & 3 & 544 & 39 & 0.014 & 0.006 & 0.047 & 0.006 & 0.092 & 0.003 & 0.049 & 0.009 & 126.8 & 5.2\\
HD191854 & R & 1 & 1400 & 101 & 0.054 & 0.023 & -0.043 & 0.023 & 0.065 & 0.011 & 0.069 & 0.033 & 70.9 & 13.5\\
HD191854 & WR & 1 & 770 & 60 & -0.021 & 0.011 & -0.060 & 0.011 & -0.027 & 0.005 & 0.063 & 0.015 & 35.2 & 6.8\\
HD193092 & WR & 1 & 128 & 9 & -0.004 & 0.010 & 0.069 & 0.010 & 0.067 & 0.004 & 0.069 & 0.014 & 136.5 & 5.8\\
HD193426 & V & 1 & 128 & 10 & 0.143 & 0.059 & 1.868 & 0.058 & 0.015 & 0.027 & 1.874 & 0.083 & 132.8 & 1.3\\
HD193426 & WR & 1 & 32 & 2 & -0.221 & 0.316 & 0.413 & 0.315 & -0.227 & 0.148 & 0.469 & 0.446 & 149.0 & 25.5\\
HD193487 & WR & 1 & 256 & 19 & -0.039 & 0.016 & -0.157 & 0.016 & -0.001 & 0.007 & 0.162 & 0.023 & 38.1 & 4.0\\
HD19373 & B & 1 & 256 & 19 & -0.096 & 0.006 & 0.231 & 0.006 & 0.085 & 0.003 & 0.250 & 0.008 & 146.3 & 1.0\\
HD19373 & R & 1 & 256 & 18 & -0.011 & 0.014 & 0.044 & 0.014 & -0.005 & 0.006 & 0.045 & 0.019 & 142.0 & 12.0\\
HD19373 & V & 1 & 192 & 14 & 0.004 & 0.006 & -0.099 & 0.006 & -0.017 & 0.003 & 0.100 & 0.008 & 46.1 & 2.4\\
HD19373 & WR & 2 & 320 & 24 & -0.029 & 0.004 & -0.011 & 0.004 & -0.013 & 0.002 & 0.031 & 0.005 & 10.5 & 4.7\\
HD195068 & WR & 2 & 506 & 59 & -0.009 & 0.006 & -0.019 & 0.006 & 0.077 & 0.003 & 0.021 & 0.008 & 33.0 & 10.8\\
HD196850 & R & 1 & 768 & 56 & -0.310 & 0.018 & -0.029 & 0.018 & -0.067 & 0.008 & 0.312 & 0.026 & 2.7 & 2.4\\
HD196850 & WR & 1 & 384 & 28 & -0.004 & 0.010 & -0.025 & 0.010 & 0.016 & 0.005 & 0.026 & 0.014 & 41.0 & 15.5\\
HD198149 & WR & 2 & 384 & 44 & 0.012 & 0.002 & -0.002 & 0.002 & -0.019 & 0.001 & 0.012 & 0.003 & 86.2 & 7.2\\
HD19820 & V & 1 & 170 & 14 & 3.121 & 0.041 & 3.670 & 0.040 & 0.055 & 0.019 & 4.817 & 0.057 & 114.8 & 0.3\\
HD19820 & WR & 2 & 144 & 10 & 2.613 & 0.019 & 3.027 & 0.019 & -0.019 & 0.009 & 3.998 & 0.027 & 114.6 & 0.2\\
HD198478 & R & 1 & 320 & 22 & -1.862 & 0.033 & -0.425 & 0.033 & 0.023 & 0.015 & 1.910 & 0.046 & 6.4 & 0.7\\
HD198478 & V & 4 & 596 & 79 & -2.638 & 0.012 & -0.089 & 0.012 & -0.021 & 0.006 & 2.639 & 0.017 & 1.0 & 0.2\\
HD198478 & WR & 8 & 1519 & 117 & -2.027 & 0.004 & 0.028 & 0.004 & -0.031 & 0.002 & 2.027 & 0.006 & 179.6 & 0.1\\
HD199100 & WR & 2 & 832 & 94 & 0.138 & 0.012 & -0.130 & 0.012 & 0.002 & 0.005 & 0.190 & 0.017 & 68.3 & 2.5\\
HD200077 & WR & 4 & 825 & 59 & -0.037 & 0.006 & 0.054 & 0.006 & 0.053 & 0.003 & 0.066 & 0.008 & 152.2 & 3.4\\
HD20630 & R & 1 & 476 & 40 & 0.021 & 0.020 & -0.121 & 0.020 & 0.077 & 0.009 & 0.123 & 0.028 & 49.8 & 6.6\\
HD20630 & WR & 1 & 128 & 9 & -0.072 & 0.006 & -0.028 & 0.006 & 0.037 & 0.003 & 0.078 & 0.009 & 10.7 & 3.4\\
HD210027 & B & 1 & 192 & 15 & 0.040 & 0.007 & 0.216 & 0.007 & 0.109 & 0.003 & 0.220 & 0.009 & 129.7 & 1.2\\
HD210027 & R & 1 & 320 & 25 & 0.074 & 0.012 & -0.037 & 0.012 & 0.050 & 0.005 & 0.083 & 0.017 & 76.7 & 5.8\\
HD210027 & V & 2 & 318 & 23 & -0.030 & 0.005 & -0.009 & 0.005 & -0.053 & 0.002 & 0.032 & 0.007 & 8.6 & 6.3\\
HD210027 & WR & 7 & 1116 & 84 & -0.002 & 0.002 & -0.034 & 0.002 & 0.019 & 0.001 & 0.034 & 0.003 & 43.1 & 2.4\\
HD212311 & R & 1 & 192 & 29 & 0.159 & 0.078 & 0.160 & 0.078 & -0.121 & 0.036 & 0.225 & 0.110 & 112.6 & 13.8\\
HD212311 & V & 1 & 576 & 43 & -0.011 & 0.018 & -0.132 & 0.018 & -0.166 & 0.008 & 0.133 & 0.025 & 42.5 & 5.5\\
HD21447 & WR & 2 & 376 & 30 & -0.242 & 0.006 & 0.020 & 0.006 & -0.015 & 0.003 & 0.243 & 0.008 & 177.7 & 1.0\\
HD214923 & B & 1 & 256 & 24 & -0.009 & 0.004 & 0.001 & 0.004 & 0.002 & 0.002 & 0.009 & 0.005 & 175.9 & 15.8\\
HD214923 & R & 2 & 768 & 58 & -0.022 & 0.009 & 0.016 & 0.009 & -0.033 & 0.004 & 0.027 & 0.013 & 162.2 & 13.1\\
HD214923 & V & 1 & 128 & 9 & -0.061 & 0.012 & -0.007 & 0.012 & -0.139 & 0.005 & 0.061 & 0.017 & 3.3 & 8.1\\
HD214923 & WR & 5 & 564 & 50 & -0.029 & 0.003 & -0.002 & 0.003 & 0.002 & 0.001 & 0.029 & 0.004 & 1.6 & 3.6\\
HD216228 & WR & 2 & 608 & 51 & -0.003 & 0.005 & 0.003 & 0.005 & -0.045 & 0.002 & 0.004 & 0.007 & 158.5 & 44.4\\
HD216411 & B & 1 & 60 & 11 & -0.031 & 0.318 & -2.482 & 0.318 & -0.094 & 0.146 & 2.483 & 0.450 & 44.6 & 5.2\\
HD216411 & V & 1 & 40 & 6 & 0.047 & 0.402 & -2.841 & 0.402 & -0.016 & 0.184 & 2.842 & 0.569 & 45.5 & 5.7\\
HD216411 & WR & 1 & 272 & 20 & -0.444 & 0.019 & -2.723 & 0.019 & -0.032 & 0.009 & 2.759 & 0.027 & 40.4 & 0.3\\
HD21770 & R & 1 & 640 & 43 & 0.053 & 0.015 & -0.026 & 0.015 & -0.027 & 0.007 & 0.059 & 0.022 & 76.8 & 10.4\\
HD222107 & R & 1 & 192 & 13 & -0.072 & 0.113 & 0.339 & 0.113 & -0.011 & 0.052 & 0.346 & 0.159 & 141.0 & 13.0\\
HD222107 & WR & 1 & 740 & 68 & -0.176 & 0.015 & -0.052 & 0.015 & -0.031 & 0.007 & 0.183 & 0.022 & 8.2 & 3.4\\
HD23512 & R & 1 & 128 & 9 & 1.314 & 0.138 & 2.307 & 0.137 & -0.001 & 0.064 & 2.655 & 0.195 & 120.2 & 2.1\\
HD23512 & WR & 2 & 384 & 28 & -1.373 & 0.021 & -1.564 & 0.021 & 0.073 & 0.010 & 2.081 & 0.030 & 24.4 & 0.4\\
HD251204 & R & 1 & 210 & 34 & -0.847 & 0.081 & 2.319 & 0.080 & -0.027 & 0.037 & 2.469 & 0.114 & 145.0 & 1.3\\
HD251204 & V & 3 & 1296 & 100 & -2.790 & 0.029 & 3.912 & 0.028 & 0.060 & 0.013 & 4.805 & 0.040 & 152.7 & 0.2\\
HD251204 & WR & 1 & 80 & 23 & -2.612 & 0.312 & 4.093 & 0.319 & -0.011 & 0.144 & 4.856 & 0.447 & 151.3 & 2.6\\
HD257971 & WR & 1 & 704 & 57 & -0.096 & 0.019 & -0.006 & 0.019 & 0.001 & 0.009 & 0.096 & 0.027 & 1.7 & 7.9\\
HD25914 & B & 1 & 22 & 4 & -0.961 & 0.065 & 4.771 & 0.067 & 0.027 & 0.030 & 4.867 & 0.093 & 140.7 & 0.5\\
HD25914 & V & 2 & 328 & 25 & -0.786 & 0.023 & 4.394 & 0.022 & 0.027 & 0.010 & 4.464 & 0.032 & 140.1 & 0.2\\
HD25914 & WR & 5 & 425 & 47 & -0.502 & 0.023 & 4.395 & 0.023 & 0.002 & 0.011 & 4.424 & 0.032 & 138.3 & 0.2\\
HD34411 & R & 1 & 576 & 41 & -0.055 & 0.013 & -0.103 & 0.013 & 0.069 & 0.006 & 0.117 & 0.019 & 30.8 & 4.6\\
HD34411 & WR & 2 & 128 & 9 & -0.001 & 0.007 & 0.037 & 0.007 & 0.072 & 0.003 & 0.037 & 0.011 & 135.5 & 8.1\\
HD39587 & WR & 2 & 256 & 18 & 0.073 & 0.005 & -0.053 & 0.005 & 0.021 & 0.002 & 0.090 & 0.008 & 72.0 & 2.4\\
            \hline
    \end{tabular}
\end{table*}

\begin{table*}
    \centering
    \caption{Continuation of the Table~\ref{tab:Table1part1}}
    \label{tab:Table1part3}
    \begin{tabular}{lccrrrrrrrrrrrr}
		\hline\hline
Object & Filter & N & NP & $\delta$T, min & q & $\sigma$q & u & $\sigma$u & v & $\sigma$v & p & $\sigma$p & $\theta$ & $\sigma\theta$\\
\hline
HD41398 & V & 1 & 320 & 27 & -2.119 & 0.026 & 1.145 & 0.026 & 0.062 & 0.012 & 2.408 & 0.037 & 165.8 & 0.4\\
HD42618 & R & 1 & 640 & 50 & 0.085 & 0.013 & -0.002 & 0.013 & -0.028 & 0.006 & 0.085 & 0.018 & 89.2 & 5.9\\
HD42618 & WR & 1 & 80 & 16 & -0.071 & 0.035 & 0.143 & 0.035 & -0.010 & 0.016 & 0.160 & 0.049 & 148.3 & 8.7\\
HD432 & R & 3 & 612 & 58 & -0.009 & 0.006 & -0.027 & 0.006 & 0.071 & 0.003 & 0.029 & 0.008 & 35.5 & 8.1\\
HD432 & WR & 7 & 752 & 67 & -0.066 & 0.002 & -0.054 & 0.002 & -0.010 & 0.001 & 0.085 & 0.003 & 19.5 & 0.9\\
HD43384 & R & 5 & 1504 & 113 & -2.012 & 0.009 & 0.854 & 0.009 & 0.004 & 0.004 & 2.186 & 0.012 & 168.5 & 0.2\\
HD43384 & WR & 6 & 1051 & 95 & -2.285 & 0.007 & 0.747 & 0.007 & 0.031 & 0.003 & 2.404 & 0.010 & 170.9 & 0.1\\
HD50019 & R & 2 & 1397 & 126 & -0.005 & 0.009 & -0.044 & 0.009 & 0.000 & 0.004 & 0.045 & 0.013 & 41.7 & 8.1\\
HD50019 & WR & 1 & 95 & 7 & 0.054 & 0.011 & 0.060 & 0.011 & 0.081 & 0.005 & 0.081 & 0.015 & 114.1 & 5.3\\
HD50973 & WR & 1 & 320 & 22 & 0.008 & 0.006 & 0.061 & 0.006 & 0.043 & 0.003 & 0.062 & 0.009 & 131.1 & 4.1\\
HD57702 & WR & 1 & 128 & 9 & -0.014 & 0.030 & 0.035 & 0.030 & -0.024 & 0.014 & 0.037 & 0.042 & 145.7 & 29.5\\
HD65583 & R & 5 & 2580 & 199 & 0.033 & 0.010 & 0.016 & 0.010 & 0.004 & 0.004 & 0.036 & 0.014 & 103.4 & 10.6\\
HD65583 & V & 3 & 864 & 66 & 0.046 & 0.013 & 0.022 & 0.014 & 0.051 & 0.006 & 0.051 & 0.019 & 102.5 & 10.6\\
HD65583 & WR & 23 & 7027 & 722 & 0.011 & 0.003 & -0.004 & 0.003 & 0.026 & 0.002 & 0.012 & 0.005 & 80.2 & 11.2\\
HD7927 & R & 1 & 224 & 16 & 1.812 & 0.006 & 0.092 & 0.006 & -0.003 & 0.003 & 1.814 & 0.008 & 91.5 & 0.1\\
HD7927 & WR & 5 & 1328 & 156 & 2.669 & 0.010 & -0.030 & 0.010 & 0.012 & 0.005 & 2.669 & 0.014 & 89.7 & 0.1\\
HD82885 & R & 3 & 2055 & 149 & 0.053 & 0.008 & -0.043 & 0.008 & 0.019 & 0.004 & 0.068 & 0.011 & 70.4 & 4.8\\
HD82885 & WR & 2 & 448 & 32 & -0.031 & 0.004 & 0.004 & 0.004 & 0.000 & 0.002 & 0.031 & 0.005 & 176.2 & 5.0\\
HD90508 & R & 2 & 959 & 74 & -0.114 & 0.016 & -0.011 & 0.016 & -0.055 & 0.007 & 0.114 & 0.022 & 2.8 & 5.6\\
HD90508 & WR & 3 & 896 & 63 & -0.026 & 0.005 & 0.014 & 0.005 & 0.005 & 0.002 & 0.030 & 0.007 & 166.1 & 7.1\\
HD9407 & V & 1 & 80 & 17 & -0.215 & 0.069 & 0.085 & 0.069 & 0.133 & 0.032 & 0.231 & 0.097 & 169.2 & 11.9\\
HD9407 & WR & 1 & 262 & 23 & 0.133 & 0.013 & -0.043 & 0.013 & -0.033 & 0.006 & 0.140 & 0.019 & 81.0 & 3.8\\
HD95418 & R & 1 & 192 & 14 & -0.089 & 0.018 & -0.079 & 0.018 & 0.114 & 0.008 & 0.119 & 0.025 & 20.8 & 6.0\\
HD95418 & WR & 1 & 128 & 9 & 0.000 & 0.012 & 0.174 & 0.012 & 0.032 & 0.005 & 0.174 & 0.018 & 135.0 & 2.9\\
HD98281 & WR & 1 & 384 & 29 & 0.002 & 0.012 & -0.012 & 0.012 & -0.002 & 0.005 & 0.012 & 0.017 & 49.1 & 34.7\\
HILTNER 960 & V & 2 & 512 & 39 & 0.938 & 0.044 & -5.314 & 0.045 & -0.024 & 0.021 & 5.396 & 0.063 & 50.0 & 0.3\\
HILTNER 960 & WR & 3 & 435 & 57 & 1.731 & 0.024 & -5.155 & 0.024 & 0.023 & 0.011 & 5.438 & 0.034 & 54.3 & 0.2\\
\hline
Object & Filter & N & NP & $\delta$T, min & Q & $\sigma$Q & U & $\sigma$U & V & $\sigma$V & P & $\sigma$P & $\theta$ & $\sigma\theta$\\
\hline
GD319 & V & 1 & 584 & 48 & 0.794 & 0.118 & 0.162 & 0.119 & 0.477 & 0.054 & 0.811 & 0.167 & 95.8 & 5.9\\
GD319 & WR & 3 & 2700 & 227 & -0.133 & 0.033 & 0.111 & 0.033 & -0.019 & 0.015 & 0.173 & 0.046 & 160.0 & 7.6\\
HD154445 & V & 1 & 128 & 9 & 3.827 & 0.072 & 0.355 & 0.074 & -0.015 & 0.032 & 3.844 & 0.104 & 92.7 & 0.8\\
HD154445 & WR & 7 & 1023 & 82 & 2.730 & 0.006 & -0.106 & 0.006 & 0.055 & 0.003 & 2.732 & 0.008 & 88.9 & 0.1\\
HD155197 & R & 1 & 640 & 111 & 3.651 & 0.093 & 1.708 & 0.092 & -0.077 & 0.043 & 4.031 & 0.131 & 102.5 & 0.9\\
HD155197 & V & 5 & 1091 & 112 & 4.089 & 0.023 & 1.981 & 0.022 & 0.025 & 0.011 & 4.543 & 0.032 & 102.9 & 0.2\\
HD155197 & WR & 5 & 899 & 74 & 2.355 & 0.018 & 1.689 & 0.018 & 0.021 & 0.008 & 2.898 & 0.025 & 107.8 & 0.3\\
HD183143 & B & 3 & 902 & 73 & -5.926 & 0.025 & 0.255 & 0.024 & 0.091 & 0.011 & 5.932 & 0.035 & 178.8 & 0.2\\
HD183143 & R & 4 & 808 & 72 & -5.632 & 0.019 & 0.581 & 0.019 & 0.042 & 0.009 & 5.662 & 0.027 & 177.1 & 0.1\\
HD183143 & V & 4 & 689 & 60 & -5.809 & 0.014 & 0.425 & 0.014 & 0.051 & 0.006 & 5.824 & 0.020 & 177.9 & 0.1\\
HD183143 & WR & 5 & 716 & 72 & -5.667 & 0.008 & -0.102 & 0.008 & 0.084 & 0.004 & 5.668 & 0.012 & 0.5 & 0.1\\
HD204827 & B & 1 & 20 & 3 & -0.107 & 0.383 & 0.483 & 0.383 & -0.032 & 0.177 & 0.495 & 0.542 & 141.2 & 28.7\\
HD204827 & R & 1 & 48 & 10 & -0.210 & 0.196 & -3.741 & 0.200 & -0.309 & 0.092 & 3.747 & 0.280 & 43.4 & 2.1\\
HD204827 & V & 4 & 598 & 59 & 2.619 & 0.016 & -4.944 & 0.016 & 0.012 & 0.008 & 5.595 & 0.023 & 59.0 & 0.1\\
HD204827 & WR & 9 & 1241 & 131 & 1.721 & 0.009 & -3.412 & 0.009 & -0.004 & 0.004 & 3.822 & 0.012 & 58.4 & 0.1\\
HD212311 & WR & 5 & 1805 & 165 & 0.579 & 0.008 & 0.113 & 0.008 & -0.012 & 0.004 & 0.590 & 0.012 & 95.5 & 0.6\\
HD21291 & WR & 2 & 194 & 13 & 1.052 & 0.007 & 0.298 & 0.007 & -0.033 & 0.003 & 1.093 & 0.009 & 97.9 & 0.2\\
HD236928 & V & 3 & 860 & 51 & 6.755 & 0.023 & 1.737 & 0.024 & -0.092 & 0.011 & 6.975 & 0.033 & 97.2 & 0.1\\
HD236928 & WR & 5 & 912 & 87 & 2.585 & 0.018 & 1.364 & 0.018 & -0.039 & 0.008 & 2.923 & 0.025 & 103.9 & 0.2\\
HD236954 & WR & 1 & 320 & 24 & 5.489 & 0.181 & 1.882 & 0.179 & -0.122 & 0.085 & 5.803 & 0.255 & 99.5 & 1.3\\
HD331891 & R & 2 & 544 & 11 & 0.618 & 0.082 & 0.307 & 0.082 & 0.070 & 0.038 & 0.689 & 0.116 & 103.2 & 4.8\\
HD331891 & V & 5 & 1791 & 81 & 0.186 & 0.017 & -0.116 & 0.017 & 0.035 & 0.008 & 0.219 & 0.023 & 74.1 & 3.1\\
HD331891 & WR & 7 & 2113 & 139 & 0.092 & 0.011 & 0.005 & 0.011 & -0.012 & 0.005 & 0.092 & 0.016 & 91.7 & 5.0\\
\hline
Object & Filter & N & NP & $\delta$T, min & Q & $\sigma$Q & U & $\sigma$U & V & $\sigma$V & P & $\sigma$P & $\theta$ & $\sigma\theta$\\
\hline
HD14489 & R & 4 & 1141 & 90 & 0.030 & 0.009 & -0.014 & 0.009 & 0.026 & 0.004 & 0.034 & 0.012 & 77.5 & 10.2\\
HD14489 & WR & 1 & 256 & 20 & 0.113 & 0.027 & 0.365 & 0.027 & 0.018 & 0.012 & 0.383 & 0.038 & 126.4 & 2.9\\
HD177463 & R & 1 & 634 & 45 & 0.068 & 0.016 & -0.037 & 0.016 & -0.010 & 0.007 & 0.077 & 0.022 & 75.7 & 8.1\\
HD177463 & WR & 1 & 246 & 22 & 0.024 & 0.007 & 0.034 & 0.007 & 0.132 & 0.003 & 0.042 & 0.010 & 117.4 & 7.1\\
HD283812 & R & 1 & 100 & 12 & -1.210 & 0.067 & -3.816 & 0.069 & -0.422 & 0.032 & 4.003 & 0.096 & 36.2 & 0.7\\
HD283812 & V & 5 & 2170 & 108 & -2.506 & 0.021 & -5.880 & 0.021 & 0.058 & 0.010 & 6.391 & 0.030 & 33.5 & 0.1\\
HD283812 & WR & 5 & 976 & 59 & -2.632 & 0.022 & -5.628 & 0.023 & 0.047 & 0.010 & 6.213 & 0.031 & 32.5 & 0.1\\
HD4768 & R & 1 & 736 & 55 & 1.741 & 0.028 & -0.833 & 0.028 & -0.112 & 0.013 & 1.930 & 0.039 & 77.2 & 0.6\\
HD161056 & WR & 2 & 448 & 36 & 2.633 & 0.011 & 2.546 & 0.012 & 0.150 & 0.005 & 3.662 & 0.016 & 112.0 & 0.1\\
HD147084 & WR & 1 & 128 & 10 & -1.092 & 0.008 & 2.874 & 0.008 & 0.111 & 0.004 & 3.074 & 0.012 & 145.4 & 0.1\\
HD34282 & WR & 1 & 544 & 54 & 1.004 & 0.171 & -2.486 & 0.325 & 0.946 & 0.147 & 2.681 & 0.305 & 56.0 & 1.9\\
HD163993 & WR & 2 & 192 & 15 & -0.013 & 0.005 & 0.045 & 0.005 & 0.133 & 0.002 & 0.047 & 0.008 & 143.1 & 4.6\\
HD90839 & R & 1 & 271 & 20 & -0.007 & 0.029 & -0.098 & 0.029 & 0.079 & 0.013 & 0.099 & 0.040 & 42.9 & 11.6\\
HD90839 & WR & 1 & 256 & 19 & 0.011 & 0.005 & 0.044 & 0.005 & -0.048 & 0.002 & 0.045 & 0.007 & 128.2 & 4.3\\
            \hline
    \end{tabular}
\end{table*}


\section{Linear and Circular Polarization}
\label{sec:LPCP}

\begin{figure*}
    \includegraphics[width=\linewidth,trim={0cm 4.65cm 0cm 5.05cm},clip]{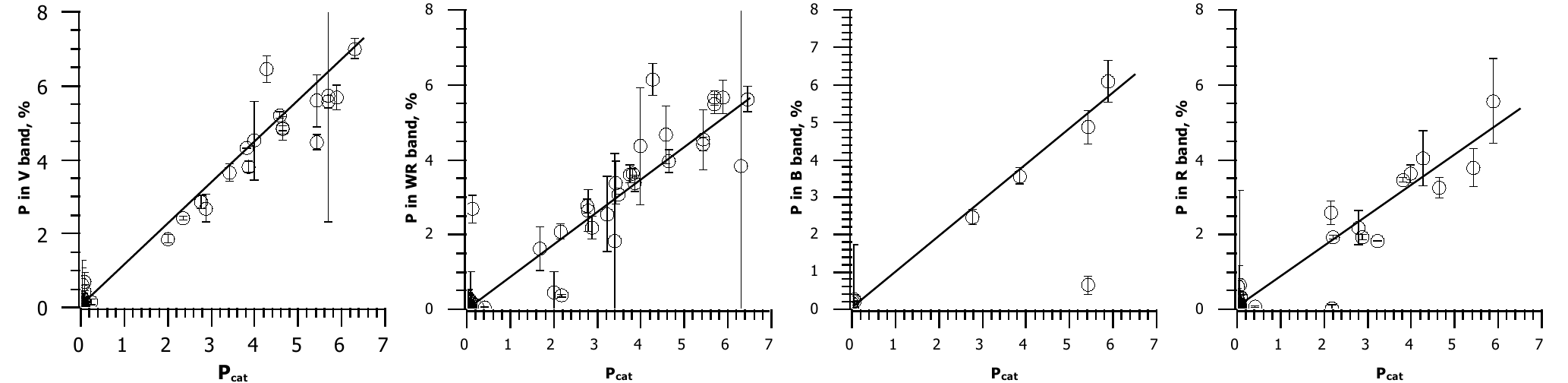}
        \caption{Dependence of the linear polarization degree for individual standards compiled for each band (from the left to the right B, V, WR, R) obtained in this work versus \citet{2000AJ....119..923H} and corresponding linear fits.}
        \label{fig:depponp}
\end{figure*}

Fig.~\ref{fig:depponp} shows a comparison of the mean weighted linear polarization degree $p$, obtained in this work, with measurements from \citet{2000AJ....119..923H} for each band. These charts contain all objects from the Table~\ref{tab:Table1part1}.

\begin{figure}
    \includegraphics[width=\linewidth]{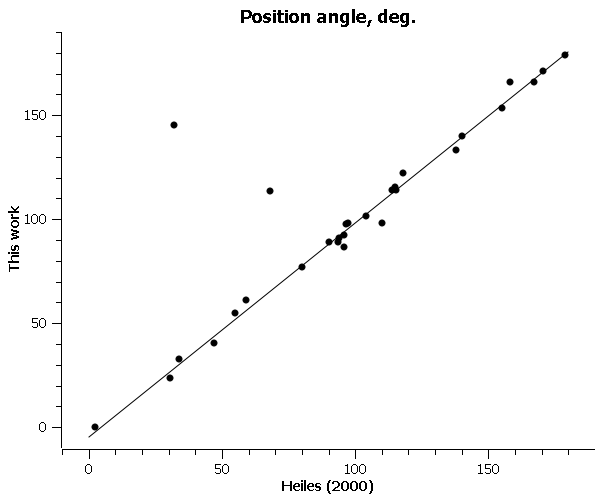}
        \caption{Dependence of the position angle for individual standards determined in this work versus \citet{2000AJ....119..923H} and corresponding linear fit. Error bars are not presented since X axis data are compiled from different sources, some points have large Y axis formal error estimate despite the data is well consistent. Two outlying points have the error estimate of the same order as other points.}
        \label{fig:depposangle}
\end{figure}

Fig.~\ref{fig:depposangle} shows a similar comparison of the position angle $\theta$ for those objects where it is determined in both sources. The angle, ranging from 0 to 180 degrees, means that values near 0 and values near 180 are the same.

A good agreement
allows us to compile all the measurements together regardless of the used filter and a season when the data were obtained for majority of tasks.

As one may see in the Table~\ref{tab:Table1part1}, for most objects the value of v, which corresponds to circular polarization, is not significant (i.e. less then 3 of its error estimate). Only 10 standard stars have statistically significant mean value of v, and 6 deserve special attention. The disagreement between linear polarization degrees from pair of sources for some standards  up to few percent may happen due to different bandwidth, different filter, variability of the source, instability on interstellar clouds or other reasons. Objects, that have linear polarization or position angle significantly different from the catalog values, are also described below.

\textbf{HD21447} shows polarization shows polarization degree $0.243(8)\%$ that is much higher then from \citet{2000AJ....119..923H}. \citet{2015MNRAS.450.1770D} also obtained higher value of $0.13(4)\%$.

The polarization degree for \textbf{HD154445} obtained in V is higher then in WR band ($3.84(10)\%$ and $2.732(8)\%$ respectively). It qualitatively confirms results obtained by different authors in V and R bands e.g. \citet{1992AJ....104.1563S} and \citet{2013NewA...20...52P}.  \citet{2007ASPC..364..529B} mentioned that this object is variable because its difference
is almost twice the limit set from the statistics.

\textbf{HD14489} has the mean value of $P = 0.07(19)\%$ for all data ($0.03(1)$ for R and $0.38(4)$ for WR band). At the same time, previous authors published the value of $2.17(20)\%$. The position angle values for WR are in a good agreement.

\textbf{HD177463} shows the linear polarization degree of $0.055(59)\%$, averaged by 2 time series obtained in different years and different bands (WR and R), where individual values are in a good agreement with each other. At the same time, the catalog value is $0.42(4)\%$.
    
For \textbf{HD283812} we obtained the polarization degree of $6.39(40)\%$, that is significantly different from the published $4.27(12)\%$. We obtained 11 time series for this object in different bands (V, R, WR) during 4 consequent years and the mean values for each set were higher then $6\%$, except one set that shows $4.04(70)\%$ that is shorter than others and show higher error estimate due to bad atmospheric conditions. Thus this set has lower weight and is not a reason to suspect the variability of this object.

For \textbf{HD4768} we obtained only 1 run, that shows the linear polarization degree $1.93(4)\%$, that is statistically different from published $2.21(20)\%$. At the same time, this run is characterized by statistically significant circular polarization of $-0.112(13)\%$.

\textbf{HD161056} was observed during 2 nights, in 2013 and 2017 in WR. The results of both time series are consistent. The degree of linear polarization is consistent with published by \citet{2000AJ....119..923H}, but the position angle is different: $68.2\pm0.8$ (Heiles \citeyear{2000AJ....119..923H}) and $112.0\pm0.1$ degrees (this work). This is an important issue, since this object is used as highly polarized standard for determination of position angle zero point by different authors. In both runs we see statistically significant low positive circular polarization about $0.15\%$. \citet{2007ASPC..364..529B} classified this object as marginal (possible variable).

\textbf{HD147084} shows similar results. We observed it during one night and obtained a linear polarization degree close to the published values, but with different polarization angle: $145.4(1)$ degrees (this work), $32.1(3)$ (Heiles \citeyear{2000AJ....119..923H}), $36.1$ in \citet{2018A&A...619A...9S}. \citet{2019A&A...622A.126P} obtained few individual values ranged in $55.14 - 81.63$ degrees in their instrumental system during consequent nights, that corresponds to approximately $8 ? 32$ degrees referred to equatorial coordinates. Their latest night $2455674.12$ has larger error estimate for p in B band, but the same in V band. It may argue for the same quality of data in V, but presence of the process which influences the polarization angle, not the degree. During our run this object showed circular polarization of $0.111(4)\%$. This object was supposed to be low-amplitude heavily-reddened star by \citet{1967ApJ...147.1003G}, but was not studied enough, so must not be used as a polarization standard.

\textbf{HD34282} is a UXOR+DSCT variable star, which shows variable polarization. We obtained $V = 0.95(15)\%$ and $P = 2.68(0.31)\%$. It is listed in catalogue by Heiles \citeyear{2000AJ....119..923H} with the value of $P = 0.13(4)\%$, but must not be used as a polarization standard.

\textbf{HD163993} was observed during 2 nights in consequent months of 2013. Both runs shows similar, relatively high and statistically significant circular polarization of $0.099(8)\%$ and $0.133(4)\%$, respectively.

\textbf{HD90839} was observed during 2 nights in 2013 and 2014. The first run is characterized by small, but statistically significant circular and linear polarizations. The second run shows only linear polarization of $0.136(44)\%$. This object is known as a star with a debris disc (Vandeportal \citeyear{2019MNRAS.483.3510V}) and a high proper-motion star (Simbad). Thus, polarized emission and the variability of the polarization could happen in this type of objects.


\section{Discussion}
\label{sec:Discussion}

\subsection{Polarimetric system}
\label{sec:polsystem}

We compared linear polarization degree p (percents) for common stars in the pairs of the data sources: \citet{2000AJ....119..923H} versus our data. The data is well agreed, but we found that the linear fit depends on the band, thus we calculated weighted linear fits in the forms of $y = k(x-x_0)+c$ and $y = kx+b$ for data obtained in each photometric filter independently. Here $x_0$ is the mean value of the x axis, that are published values of the linear polarization degree. The coefficients of
the first one is presented in the Table~\ref{tab:polkoef}.

We noticed that these fits are statistically different and this approach may be used not only for polarization scale calibration, but also for converting the results obtained using different photometric systems to one system similarly to how it is used to be in photometry.

\begin{table}
    \centering
    \caption{The coefficients of the weighted linear fit to dependence of linear polarization degree p for common stars in the pairs of the data sources.}
    \label{tab:polkoef}
    \begin{tabular}{lcccccr} 
		\hline\hline
Band & k & $\sigma k$ & c & $\sigma c$ & $x_0$ & $\chi^2$ \\
            \hline
B & 0.9564 & 0.1272 & 1.6240 & 0.2065 & 1.6460 & 112 \\
R & 0.8127 & 0.0238 & 0.8440 & 0.0237 & 0.9493 & 140 \\
U & 1.0215 & 0.1504 & 1.7339 & 0.2599 & 1.4258 & 42 \\
V & 1.1115 & 0.0079 & 2.2230 & 0.0152 & 1.9635 & 214 \\
WR & 0.8737 & 0.0041 & 0.9862 & 0.0058 & 1.0961 & 214 \\
            \hline
    \end{tabular}
\end{table}

\subsection{The Variability of Standard Stars}
\label{sec:stvariability}

We built time series of each Stokes parameter of each standard star by each band separately for full time interval and analysed them using well known numerical parameters characterizing the degree of variability (so called variability indices). We chose scatter-based indices, that we used earlier for variability detection on photometric data (see \citet{2017AASP....7....3B}, \citet{2019OEJV..197...61B}): standard deviation, $\chi^2$ test, Median absolute deviation, Robust median statistic and Normalized excess variance. Also we used Welch test (see \citet{1986VA.....29...27C}) for all time series. We marked all time series where at least one variability detection index shows an outstanding value. All stars, that have at least one time series marked few times were moved to the penultimate section of the Table~\ref{tab:Table1part1}. These stars are suspected to exhibit variability of at least one Stokes parameter, so they should not be used as polarized standard stars without further investigation.


\section{Conclusions}
\begin{enumerate}

\item
We re-determined values of Stokes parameters for 98 standard stars (both non-polarized and highly polarized) and presented them along with the polarization degree p in the tables. This catalogue is not compilative, but obtained using the same instrument and technique during large time interval. It will be used for our future research and it may be used by other authors.

\item
We implemented the least squares approach for determination of the Stokes parameters. It allowed us to obtain results with better accuracy.

\item
Due to the dependence of polarization degree on the bandwidth, the statistical dependencies may be used for calibration and converting the results obtained using different photopolarimetric systems to one system similarly to how it is used to be in photometry.

\item
The determination of the position angle zero point should be done using at least 3 highly polarized standards to exclude possible errors. One of chosen standard stars may have a position angle which varies with wavelength (which suggests that there are multiple polarization components) or time (polarization component variability). Passing of polarized interstellar clouds may also temporary effect the observable position angle. Despite this is an obvious conclusion, some authors use only one or two standards.

\item
The same is true for zero-polarization standards.

\item
Usage of insufficiently explored standards is possible only for determination of the correct coordinate quarter or clarification of the mean values of instrumental polarization together with well known reliable standards. This approach will also increase the number of well known standards.

\item
It is also not desirable to use only agglomeration catalogs.

\item
Although Stokes parameters may have different statistical distribution, we adopted
selected scatter-based indices widely used for analysis of photometric data and compared with Welch test. We obtained similar results that means these scatter-based indices may be used for detection of polarization variability and it was confirmed on large array of uniform data.
We report suspicious or variable stars that should no longer be used as polarized standards to obtain precise results.


\end{enumerate}


%
%
\bibliographystyle{aa} 
\bibliography{paper1bibaa} 

\begin{thebibliography}{}
\makeatletter
\relax
\def\mn@urlcharsother{\let\do\@makeother \do\$\do\&\do\#\do\^\do\_\do\%\do\~}
\def\mn@doi{\begingroup\mn@urlcharsother \@ifnextchar [ {\mn@doi@}
  {\mn@doi@[]}}
\def\mn@doi@[#1]#2{\def\@tempa{#1}\ifx\@tempa\@empty \href
  {http://dx.doi.org/#2} {doi:#2}\else \href {http://dx.doi.org/#2} {#1}\fi
  \endgroup}
\def\mn@eprint#1#2{\mn@eprint@#1:#2::\@nil}
\def\mn@eprint@arXiv#1{\href {http://arxiv.org/abs/#1} {{\tt arXiv:#1}}}
\def\mn@eprint@dblp#1{\href {http://dblp.uni-trier.de/rec/bibtex/#1.xml}
  {dblp:#1}}
\def\mn@eprint@#1:#2:#3:#4\@nil{\def\@tempa {#1}\def\@tempb {#2}\def\@tempc
  {#3}\ifx \@tempc \@empty \let \@tempc \@tempb \let \@tempb \@tempa \fi \ifx
  \@tempb \@empty \def\@tempb {arXiv}\fi \@ifundefined
  {mn@eprint@\@tempb}{\@tempb:\@tempc}{\expandafter \expandafter \csname
  mn@eprint@\@tempb\endcsname \expandafter{\@tempc}}}

\bibitem[\protect\citeauthoryear{{Bastien}, {Vernet}, {Drissen}, {M{\'e}nard},
  {Moffat}, {Robert}  \& {St-Louis}}{{Bastien}
  et~al.}{2007}]{2007ASPC..364..529B}
{Bastien} P.,  {Vernet} E.,  {Drissen} L.,  {M{\'e}nard} F.,  {Moffat}
  A.~F.~J.,  {Robert} C.,   {St-Louis} N.,  2007, in {Sterken} C.,  ed.,
  Astronomical Society of the Pacific Conference Series Vol. 364, The Future of
  Photometric, Spectrophotometric and Polarimetric Standardization. p.~529

\bibitem[\protect\citeauthoryear{{Breus}}{{Breus}}{2007}]{2007OAP....20...32B}
{Breus} V.~V.,  2007, Odessa Astronomical Publications, \href
  {https://ui.adsabs.harvard.edu/abs/2007OAP....20...32B} {20, 32}

\bibitem[\protect\citeauthoryear{{Breus}}{{Breus}}{2017}]{2017AASP....7....3B}
{Breus} V.~V.,  2017, \mn@doi [Advances in Astronomy and Space Physics]
  {10.17721/2227-1481.7.3-5}, \href
  {https://ui.adsabs.harvard.edu/abs/2017AASP....7....3B} {7, 3}

\bibitem[\protect\citeauthoryear{{Breus}}{{Breus}}{2019}]{2019OEJV..197...61B}
{Breus} V.~V.,  2019, Open European Journal on Variable Stars, \href
  {https://ui.adsabs.harvard.edu/abs/2019OEJV..197...61B} {197, 61}

\bibitem[\protect\citeauthoryear{{Breus}, {Andronov}, {Kolesnikov}  \&
  {Shakhovskoy}}{{Breus} et~al.}{2007}]{2007A&AT...26..241B}
{Breus} V.~V.,  {Andronov} I.~L.,  {Kolesnikov} S.~V.,   {Shakhovskoy} N.~M.,
  2007, \mn@doi [Astronomical and Astrophysical Transactions]
  {10.1080/10556790701243225}, \href
  {https://ui.adsabs.harvard.edu/abs/2007A&AT...26..241B} {26, 241}

\bibitem[\protect\citeauthoryear{{Breus} et~al.,}{{Breus}
  et~al.}{2013}]{2013JPhSt..17.3901B}
{Breus} V.~V.,  et~al., 2013, Journal of Physical Studies, \href
  {https://ui.adsabs.harvard.edu/abs/2013JPhSt..17.3901B} {17, 3902}

\bibitem[\protect\citeauthoryear{{Clarke} \& {Stewart}}{{Clarke} \&
  {Stewart}}{1986}]{1986VA.....29...27C}
{Clarke} D.,  {Stewart} B.~G.,  1986, \mn@doi [Vistas in Astronomy]
  {10.1016/0083-6656(86)90013-9}, \href
  {https://ui.adsabs.harvard.edu/abs/1986VA.....29...27C} {29, 27}

\bibitem[\protect\citeauthoryear{{Deb Roy}, {Halder}, {Das}  \& {Medhi}}{{Deb
  Roy} et~al.}{2015}]{2015MNRAS.450.1770D}
{Deb Roy} P.,  {Halder} P.,  {Das} H.~S.,   {Medhi} B.~J.,  2015, \mn@doi
  [\mnras] {10.1093/mnras/stv707}, \href
  {https://ui.adsabs.harvard.edu/abs/2015MNRAS.450.1770D} {450, 1770}

\bibitem[\protect\citeauthoryear{{Efimov}}{{Efimov}}{1981}]{1981IzKry..63..118E}
{Efimov} I.~S.,  1981, Izvestiya Ordena Trudovogo Krasnogo Znameni Krymskoj
  Astrofizicheskoj Observatorii, \href
  {https://ui.adsabs.harvard.edu/abs/1981IzKry..63..118E} {63, 118}

\bibitem[\protect\citeauthoryear{{Garrison}}{{Garrison}}{1967}]{1967ApJ...147.1003G}
{Garrison} R.~F.,  1967, \mn@doi [\apj] {10.1086/149090}, \href
  {https://ui.adsabs.harvard.edu/abs/1967ApJ...147.1003G} {147, 1003}

\bibitem[\protect\citeauthoryear{{Heiles}}{{Heiles}}{2000}]{2000AJ....119..923H}
{Heiles} C.,  2000, \mn@doi [\aj] {10.1086/301236}, \href
  {https://ui.adsabs.harvard.edu/abs/2000AJ....119..923H} {119, 923}

\bibitem[\protect\citeauthoryear{{Kolesnikov}}{{Kolesnikov}}{2019a}]{2019JPhSt..23.3901K}
{Kolesnikov} S.~V.,  2019a, \mn@doi [Journal of Physical Studies]
  {10.30970/jps.23.3901}, \href
  {https://ui.adsabs.harvard.edu/abs/2019JPhSt..23.3901K} {23, 3901}

\bibitem[\protect\citeauthoryear{{Kolesnikov}}{{Kolesnikov}}{2019b}]{2019JPhSt..23.4901K}
{Kolesnikov} S.~V.,  2019b, \mn@doi [Journal of Physical Studies]
  {10.30970/jps.23.4901}, \href
  {https://ui.adsabs.harvard.edu/abs/2019JPhSt..23.4901K} {23, 4901}

\bibitem[\protect\citeauthoryear{{Kolesnikov}, {Breus}, {Kiselev}  \&
  {Andronov}}{{Kolesnikov} et~al.}{2016}]{2016OAP....29...74K}
{Kolesnikov} S.~V.,  {Breus} V.~V.,  {Kiselev} N.~N.,   {Andronov} I.~L.,
  2016, \mn@doi [Odessa Astronomical Publications]
  {10.18524/1810-4215.2016.29.85015}, \href
  {https://ui.adsabs.harvard.edu/abs/2016OAP....29...74K} {29, 74}

\bibitem[\protect\citeauthoryear{{Kucherov}}{{Kucherov}}{1986}]{1986KFNT....2...82K}
{Kucherov} V.~A.,  1986, Kinematika i Fizika Nebesnykh Tel, \href
  {https://ui.adsabs.harvard.edu/abs/1986KFNT....2...82K} {2, 82}

\bibitem[\protect\citeauthoryear{{Leroy}}{{Leroy}}{1993}]{1993A&AS..101..551L}
{Leroy} J.~L.,  1993, \aaps, \href
  {https://ui.adsabs.harvard.edu/abs/1993A&AS..101..551L} {101, 551}

\bibitem[\protect\citeauthoryear{{Mishchenko} et~al.,}{{Mishchenko}
  et~al.}{2010}]{2010Mishchenko}
{Mishchenko} M.~I.,  et~al., 2010, {Polarimetric Remote Sensing of Solar System
  Objects}

\bibitem[\protect\citeauthoryear{{Pavana}, {Anche}, {Anupama}, {Ramaprakash}
  \& {Selvakumar}}{{Pavana} et~al.}{2019}]{2019A&A...622A.126P}
{Pavana} M.,  {Anche} R.~M.,  {Anupama} G.~C.,  {Ramaprakash} A.~N.,
  {Selvakumar} G.,  2019, \mn@doi [\aap] {10.1051/0004-6361/201833728}, \href
  {https://ui.adsabs.harvard.edu/abs/2019A&A...622A.126P} {622, A126}

\bibitem[\protect\citeauthoryear{{Prasad}, {Pandey}, {Patel}  \&
  {Srivastava}}{{Prasad} et~al.}{2013}]{2013NewA...20...52P}
{Prasad} V.,  {Pandey} J.~C.,  {Patel} M.~K.,   {Srivastava} D.~C.,  2013,
  \mn@doi [\na] {10.1016/j.newast.2012.10.001}, \href
  {https://ui.adsabs.harvard.edu/abs/2013NewA...20...52P} {20, 52}

\bibitem[\protect\citeauthoryear{{Rosenbush}, {Kiselev}, {Antoniuk}  \&
  {Kolesnikov}}{{Rosenbush} et~al.}{2008}]{2008LPICo1405.8393R}
{Rosenbush} V.,  {Kiselev} N.,  {Antoniuk} K.,   {Kolesnikov} S.,  2008, in
  Asteroids, Comets, Meteors 2008. p.~8393

\bibitem[\protect\citeauthoryear{{Schmid} et~al.,}{{Schmid}
  et~al.}{2018}]{2018A&A...619A...9S}
{Schmid} H.~M.,  et~al., 2018, \mn@doi [\aap] {10.1051/0004-6361/201833620},
  \href {https://ui.adsabs.harvard.edu/abs/2018A&A...619A...9S} {619, A9}

\bibitem[\protect\citeauthoryear{{Schmidt}, {Elston}  \& {Lupie}}{{Schmidt}
  et~al.}{1992}]{1992AJ....104.1563S}
{Schmidt} G.~D.,  {Elston} R.,   {Lupie} O.~L.,  1992, \mn@doi [\aj]
  {10.1086/116341}, \href
  {https://ui.adsabs.harvard.edu/abs/1992AJ....104.1563S} {104, 1563}

\bibitem[\protect\citeauthoryear{{Serkowski}}{{Serkowski}}{1974}]{1974psns.coll..135S}
{Serkowski} K.,  1974, in {Gehrels} T.,  ed., IAU Colloq. 23: Planets, Stars,
  and Nebulae: Studied with Photopolarimetry. p.~135

\bibitem[\protect\citeauthoryear{{Shakhovskoj} \& {Efimov}}{{Shakhovskoj} \&
  {Efimov}}{1972}]{1972IzKry..45...90S}
{Shakhovskoj} N.~M.,  {Efimov} Y.~S.,  1972, Izvestiya Ordena Trudovogo
  Krasnogo Znameni Krymskoj Astrofizicheskoj Observatorii, \href
  {https://ui.adsabs.harvard.edu/abs/1972IzKry..45...90S} {45, 90}

\bibitem[\protect\citeauthoryear{{Shakhovskoj}, {Andronov}, {Kolesnikov}  \&
  {Khalevin}}{{Shakhovskoj} et~al.}{1998}]{1998KPCB...14..359S}
{Shakhovskoj} N.~M.,  {Andronov} I.~L.,  {Kolesnikov} S.~V.,   {Khalevin}
  A.~V.,  1998, Kinematics and Physics of Celestial Bodies, \href
  {https://ui.adsabs.harvard.edu/abs/1998KPCB...14..359S} {14, 359}

\bibitem[\protect\citeauthoryear{{Shakhovskoy}, {Andronov}, {Kolesnikov}  \&
  {Khalevin}}{{Shakhovskoy} et~al.}{2001}]{2001IzKry..97...91S}
{Shakhovskoy} N.~M.,  {Andronov} I.~L.,  {Kolesnikov} S.~V.,   {Khalevin}
  A.~V.,  2001, Izvestiya Ordena Trudovogo Krasnogo Znameni Krymskoj
  Astrofizicheskoj Observatorii, \href
  {https://ui.adsabs.harvard.edu/abs/2001IzKry..97...91S} {97, 91}

\bibitem[\protect\citeauthoryear{{Vandeportal}, {Bastien}, {Simon}, {Augereau}
  \& {Storer}}{{Vandeportal} et~al.}{2019}]{2019MNRAS.483.3510V}
{Vandeportal} J.,  {Bastien} P.,  {Simon} A.,  {Augereau} J.-C.,   {Storer}
  {\'E}.,  2019, \mn@doi [\mnras] {10.1093/mnras/sty3060}, \href
  {https://ui.adsabs.harvard.edu/abs/2019MNRAS.483.3510V} {483, 3510}

\makeatother
\end{thebibliography}

\end{document}